\documentstyle[12pt]{article}
\input psfig.tex
\begin{document}
\centerline{\large\bf Extra Dimensions and Strong Neutrino-Nucleon
Interactions}
\centerline{\large\bf  Above $10^{19}$ eV : Breaking the GZK Barrier
}

\bigskip
\centerline{\large\bf
P. Jain$^a$, Douglas W. McKay$^b$, S. Panda$^a$ and John P. Ralston$^b$}

\bigskip
\begin{center}
$^a$ Physics Department\\
	I.I.T. Kanpur, India 208016

$^b$	Department of Physics \& Astronomy\\
	University of Kansas\\
	Lawerence, KS 66045, USA
\end{center}

\bigskip
\noindent
{\bf Abstract:} Cosmic ray events above $10^{20}$ eV are on the verge of
confronting fundamental particle physics. The neutrino is the only
candidate primary among established particles
capable of crossing 100 Mpc 
intergalactic distances unimpeded.  
The
magnitude of $\nu N$  cross sections indicated by events, plus consistency
with the Standard Model at  low-energy, point to new physics of massive
spin-2 exchange.  In models based on extra dimensions, we find that the
$\nu N$ cross section rises to typical hadronic values of between 1 and 100
mb at energies above $10^{20}$ eV.  Our calculations take into account
constraints of unitarity.  We conclude that air-showers observed with
energies above  $10^{19}$ eV are consistent with neutrino primaries and
extra-dimension models.   An {\it upper bound} of 1-10 TeV on the mass
scale at which graviton exchange becomes strong in current Kaluza-Klein
models follows.

\section{Introduction} The energy of extra-galactic proton
cosmic rays should not exceed the GZK bound \cite{GZK}.  The bound, about
$10^{19}$ eV, is based on the known interactions of nucleon primaries with
the photon background of intergalactic space.  The $GZK$ bound is
tantamount to an upper limit on cosmic ray energies, inasmuch as nuclei and
photons have lower energy cutoffs \cite{puget}.  Yet an experimental
puzzle exists, as evidence for air shower events with energies above the
GZK bound has steadily accumulated over the last 35
years \cite{linsley}.  There seem to be inadequate
sources nearby to account for such events, and the sources are almost
certainly extragalactic \cite{galaxy}.  A number of showers with energies
reliably determined to be above $10^{20}$ eV have been observed in recent
years \cite {fliesI}, deepening the puzzle.

A completely satisfactory explanation of the so-called $GZK$-violating events
is still lacking.  Models have been constructed to explain the puzzle by
invoking ``conventional'' extragalactic sources of proton $UHE$ acceleration,
as reviewed recently in Ref.\cite{bland}.  Other models introduce exotic
primaries such as magnetic monopoles \cite{mono} or exotic sources such as
unstable superheavy relic particles \cite{heavy}, or appeal to topological
defects \cite{topdef}. 
Except for monopoles all of the proposed sources must be within 50-100
Mpc to evade the GKZ bound, a requirement which is difficult to satisfy.

The $10^{20}$ eV events potentially pose a confrontation between
observation and fundamental particle physics.  Except for the neutrino,
there are no candidates among $\it{established}$ elementary particles that
could cross the requisite intergalactic distances of about 100 Mpc or more
\cite{bewax,Fargion}.  The neutrino would be a candidate
for the events, if its interaction cross section were large enough,
but this requires physics beyond the Standard Model.  The neutrino-nucleon
total cross section $\sigma_{tot}$ is the
crucial issue: Flux estimates of $UHE$ neutrinos produced by extra-galactic
sources and GZK-attenuated nucleons and nuclei vary widely, but suffice to
account for the shower rates observed.  Very significantly, there is a hint
of correlations in direction of the events, both with one another and
candidate sources \cite{Farrar98}. 
Event-pointing toward distant sources, if confirmed, would require a
neutral, long-lived primary, reducing the possibilities for practical
purposes to the neutrino plus new physics to explain the cross section.

Current understanding of the $UHE$ Standard Model $\sigma_{tot}$ is based
on small-x QCD evolution and $W^{\pm}, Z$ exchange physics \cite{q+r}. 
This physics is extremely well understood and has been directly tested up to
$s = 10^5$ GeV$^2$ with recent HERA-based updates \cite{Frich96}. 
These calculations are then extrapolated to
the region of $10^{20}$ eV primary energy, leading to cross sections in the
range $10^{-4}\ -\ 10^{-5}$ mb, far too small to explain the GZK-violating
air shower events.  The observationally
indicated cross section is completely out of reach of the extrapolations of
$W$- and $Z$- exchange mechanisms.

Since the neutrino-nucleon cross section at $10^{20}$ eV has never been
directly measured, it is quite reasonable to surmise that new physical
processes may be at work.  Total cross sections at high energies are dominated
by characteristics of the t-channel exchanges.  The growth of $\sigma_{tot}$
with energy, in turn, is directly correlated with the {\it spin} of
exchanged particles.  Exchange of new (W- or Z-like) massive {\it vector}
bosons would produce $\sigma_{tot}$ growing at the same rate as the
standard one, failing to explain the puzzle.  If the data indicates a more
rapid growth with energy, one is forced to consider higher spin, with the
next logical possibility being {\it massive spin-2} exchange.  We reiterate
that this deduction is data-driven; if data indicates (a) corrrelations
with source directions, and (b) $\sigma_{tot}$ in the $mb$ and above
range, there are
few options other than neutrinos interacting by massive spin-2 exchanges.

Recent theoretical progress has opened up the fascinating possibility of
massive spin-2, t-channel exchange in the context of large ``extra" 
dimensions \cite{ADD1}, while the fundamental scale
can be related to a string scale of order several TeV \cite{ADD2}.
 In this context the Kaluza-Klein ($KK$) excitations of the
graviton act like a tower of massive spin-2 particle exchanges .
We will show that large $UHE$ neutrino cross sections, sufficient to
generate the observed showers, are a generic feature of this developing
framework \cite {d+d}.  At the same time the new contributions to
$\sigma_{tot}$ at
energies below center of mass energy $\surd s$ of $500$ GeV
is several orders of magnitude below the Standard Model
component.  In fact, the new physics we propose to explain the puzzle of
the $GZK$-violating events is consistent with all known experimental
limits.

\section{The $\nu N$ Cross Section with Massive Spin-2 Exchange}

The low-energy, 4-Fermi interaction total cross section, $\sigma_{tot}$,
grows like $s^1$ over many decades of energy.
Perturbative unitarity implies that at an invariant $cm$ energy
$\sqrt{s}$ large compared to the exchange mass $m_{W}$, the growth rate
slows to at most a logarithmic energy dependence.  The shift from power-law
to logarithmic growth is seen to occur in the Standard Model.  There is a
second effect, that above 100
TeV the total number of targets (quark-antiquark pairs) grows roughly like
$(E_{\nu})^{0.4}$.  This fractional power, in turn, leads to a formula for
$\sigma_{tot} = 1.2 \times 10^{-5} mb (E_{\nu}/10^{18}{\rm eV})^{0.4}$ as a
reasonable approximation to the Standard Model calculation \cite{Frich96}.

Exchange of additional spin-1 bosons cannot produce faster growth with energy
than just described.  However, a massive spin-2 exchange grows quite
quickly with energy on very general grounds.  A dimensionless spin-2 field
gets its couplings from derivatives, which translate to factors of
energy.  Thus the naive cross section grows like $E^{3}_{\nu}$, in the
``low-energy'' regime.

These general features are exemplified in the Feynman rules for this regime
developed by several groups \cite{HaLyZh}.  To be consistent with the
literature, we will describe the interaction as ``graviton" exchange,
implying the standard picture of a tower of spin-2 $KK$ modes.  The parton
level $\nu$ gluon differential cross section is given by,
\begin{equation}
 {d\hat\sigma^{Gg} \over d \hat t} = {\pi\lambda^2\over 2 M_S^8}{\hat
 u\over
\hat s^2}\left[2\hat u^3 + 4\hat u^2\hat t+3 \hat u\hat t^2+\hat t^3\right]
\end{equation}
Here $M_{S}$ is the cutoff on the graviton mass, and $\lambda$ is the
effective coupling at the scale $M_{S}$ that cuts off the graviton $KK$
mode summation.  The magnitude of parameter $\lambda$ has been lumped into
the scale parameter $M_S$, hence $\lambda=\pm 1$ for our purposes 
\cite{mathews}.  In
Eqs.  (1) and (2) we take $-\hat t\ll M_S^2$, which leads to the simple
factor $1/M_{S}^8$.  This suffices for our extrapolation, but
the full $\hat t$ dependence is used in the partial wave projections to
check the unitarity constraint (discussed momentarily).  The corresponding
parton level $\nu$ quark differential cross section is given by,
\begin{equation}
 {d\hat\sigma^{Gq} \over d \hat t} = {\pi\lambda^2\over 32 M_S^8}{1\over
\hat s^2}\left[32\hat u^4 + 64\hat u^3\hat t+42 \hat u^2\hat t^2+
10\hat u\hat t^3+\hat t^4\right]
\end{equation}
We include the contribution of the two valence quarks as well as the $\bar
u$, $\bar d$ and $\bar s$ sea quarks.  The $Z$-graviton interference terms
are included with negative $\lambda$, though their contribution is very
small compared to other terms.  The negative sign gives a slight
enhancement for the final result of $\sigma_{tot}$.  Collider
physics and astrophysics constrain the effective scale $M_S$ to be above 1
TeV, with lower number of dimensions leading to stronger constraints.

\subsection{Unitarity}

The complete theory of massive $KK$ modes is not yet developed, making it
impossible to know the exact cross sections at asymptotic energies.  The
situation is analogous to the case of the 4-Fermi theory before the
Standard Model.  By observing the $s^1$ growth of $\sigma_{tot}$ it was
possible to deduce a massive vector exchange long before a consistent 
theory existed.  In much the same way, present data indicate a spin-2
exchange while the analogous complete ``standard model'' of gravitons
does not yet exist.  Unlike the electroweak case, in either the data-
driven or extra-dimensions scenario we must face a strongly interacting,
non-perturbative problem in the high energy, $s\gg M_{S}^2$.  Perturbative
unitarity breaks down as a host of new channels opens up in that regime. 
The low-energy effective theory remains an
accurate description within a particular domain of consistency.
Extrapolation of the 4-Fermi predictions to higher energies is possible by
matching the consistent, low-energy description with the asymptotic demands
of unitarity.  Similarly, we resolve the difficulties of massive graviton
exchange
in the high energy regime by matching the $\surd s < M_{S}$ predictions,
where the perturbative calculation is under control, to the $\surd s\gg
M_{S}$ non-perturbative regime.

We proceed by first evaluating the theory's partial wave amplitudes to find
the highest energy where the low energy effective theory is applicable.
Taking the case $\nu+q\rightarrow \nu+q$, and including the full $Q^2$
dependence of the propagator, we find that the unitarity bound on the $J=0$
projection of the helicity amplitude $T_{++,++}$
\cite{MaSp} gives the strongest bound.  For example, with the number
of extra dimensions $n=2$ we find $\surd s\leq 1.7 M_{S}$, while with $n=4$
we find $\surd s\leq 2.0 M_{S}$.  As mentioned
earlier, the most attractive value of $M_S$ is in the TeV range. 

The
invariant energies of the highest energy cosmic rays are approximately 1000
units of the scale $M_{S}\sim 1$ TeV, well beyond the low-energy regime.  A
phenomenological prescription consistent with unitarity is clearly
necessary to extrapolate the low energy amplitudes.   We now turn to
describing and motivating three different asymptotic forms 
that span reasonable possibilities: $log(s)$, $s^1$ and $s^2$ growth of  
$\sigma_{tot}(s)$.  There is
no guarantee a priori that any should extrapolate from low to high energy
with $M_{S} \sim 1$ TeV and produce hadronic-size cross sections at $10^{20}$
eV.  As we shall see, surprisingly, they all do!

As a first version of an extrapolation model, we use a
well known result from general features of local quantum field theory.  The
Froissart bound \cite{Froi}, reflecting the unitarity constraint on cross
sections from exchange of massive particles, dictates that $\sigma_{tot}$
grows no more rapidly than $\left({\rm log}(\hat s/M_S^2)\right)^2$.  The
bound is an asymptotic one, and strictly speaking incapable of limiting
behavior at any finite energy; moreover, the bound is probably
violated in the case of graviton exchange.  It is quite conservative to use the
Froissart bound, with its mild logarithmic growth in s, as a first test case. 
 In terms of the differential cross section, we have at high energy
\begin{equation}
 {d\hat\sigma \over d \hat t}\rightarrow {{\rm const}\over \hat t M_S^2}
 {\rm log} (\hat s/M_S^2)
\end{equation}
We then propose the following interpolating formula which reproduces Eq.
(1) in the low energy limit and Eq.  (3) in the high energy limit
\begin{eqnarray}
 {d\hat\sigma^{Gg} \over d \hat t} &=& {\pi\lambda^2\over 2 M_S^2
 (M_S^2+\hat s)^2 (M_S^2-\hat t)}{\hat u\over
\hat s^2}\left[2\hat u^3 + 4\hat u^2\hat t+3 \hat u\hat t^2+\hat
t^3\right]\cr &\times & \left[1+\xi {\rm log}(1+\hat s/M_S^2)\right]
\end{eqnarray}
The $\nu$-quark parton level cross section is similarly extrapolated to
high energies.  We have introduced the parameter $\xi$, which we will allow
to vary between 1 and 10.  It cannot be much larger than 10 since then the low
energy cross section gets modified, violating consistency.  We use these
parton-level expression to calculate
$\sigma_{tot}$.  We convolute the parton-level cross section with CTEQ 4.6
parton distribution functions, which give a continued growth in the $UHE$
regime from the small-x effect.  As pointed out previously, the cross
section can be expected to grow to ``strong interaction" magnitudes, where
parton coalescence and string effects ultimately come into play.

A different constraint for unitarizing would be $s^2$ growth.  Regge
theory would suggest $ s^2$ growth for spin-2 exchange at small, fixed $t$,
which (in fact) occurs in this theory when the $t$ values are restricted
self-consistently.  
Thus the use of $s^2$ growth makes a
comparatively mild alteration of the perturbative predictions and
follows from eikonal unitarization of Reggeized graviton
exchange \cite{muzinich}. Another unitarization procedure indicates
a linear growth in $s$  
\cite{N+S}, which represents a case intermediate to the other two.  
These cases serve to establish a fair range of
possibilities for models of unitarization.  For the $s^1$ and $s^2$ models,
the extrapolation form of Eq. (1) we choose is

\begin{equation}
 {d\hat\sigma^{Gg} \over d \hat t} = {\pi\lambda^2\over 2 M_S^{6-2p} (M_S^2+\hat s)^p
(M_S^2-\beta \hat t)}{\hat u\over
\hat s^2}\left[2\hat u^3 + 4\hat u^2\hat t+3 \hat u\hat t^2+\hat t^3\right],
\end{equation}
where $p=1, 0$ for $s^1, s^2$. 
A similar extrapolation is applied to the cross section formula for the $\nu$
quark parton case, Eq. (2).

Note that we use the detailed perturbative low energy calculations that
follow from \cite{HaLyZh} to anchor the low energy end.  As we
discuss below, we also explore the allowed range of $M_S$ values,
corresponding to a range of numbers of extra large dimensions. As far as we
know, these consistency features in the present context have not previously 
appeared in the literature.

An essential feature of spin-2 exchange, complementary to the growth of
$\sigma_{tot}$ with energy in the $UHE$ region, is {\it suppression} of
observable effects in the low energy
regime of existing data.  Turning the problem around to a ``data driven"
view, the energy dependence of the new physics must be so strong that the
millibarn-scale total cross sections at $\surd s\geq 10^3$ TeV are
suppressed well below the Standard
Model values below 1 TeV.  This also follows in our approach.

\section{Results}

Results of the calculations based on our models (Eqs. 4 and 5) are given in
Fig. 1.  For the $ln(s)$ model, shown by the dotted line,  
we find that for incident neutrino energy of order $10^{12}$
GeV, $\sigma_{tot}$ is roughly $0.5$ mb, with the effective
cutoff scale $M_S=1$ TeV and the choice $\xi = 10$. 
This cross section is remarkably consistent
with the low end of the range required to
explain the GZK violating events.  Looking at higher values of $M_{S}$, we
find that the $\sigma_{tot}$ falls so steeply as a function of $M_{S}$ that
$M_{S}\approx 1$ TeV is $\it{required}$ for this model to be a viable
explanation of cosmic ray events with $E\geq 10^{19}$ eV.

The results of the calculations with the $s^1$  model are shown in
the dash-dotted line ($\beta=0.1$) and short-dashed line ($\beta=1.0$).
 The value of $M_{S}$ is fixed at 1 TeV; $\beta$
values of 0.1 and 1.0 are chosen to illustrate that this intermediate growth
model easily produces hadronic size cross sections above the GZK cutoff.  We
find that raising the cutoff $M_{S}$  much above 2 TeV suppresses the
cross section well below the range required for the GZK events, independently
of the value of $\beta$. 

The long-dashed curve (Fig. 1) shows the case with $s^2$ growth
for values $M_{S} = 3$ TeV and $\beta = 1.0$.  The growth of
$\sigma_{tot}$ to tens of millibarns at the highest energies shows the
viability of this model even at $M_{S}$ values above 1 TeV.
 
The key feature we emphasize about Fig. 1 is
that $\sigma_{tot}$ somewhere in the range 1-100 mb is obtained at the
highest energies for reasonable parameter choices for all three 
unitarization models.  As a significant corollary, we find
that if $M_S$ is larger than 1 - 10 TeV (depending on the model), 
then the low scale gravity
models fail to increase $\sigma_{tot}$ by enough to explain the
$GZK$-violating events.  Thus the analysis serves to put
an {\it upper bound} on the cutoff mass\footnote{Our requirement is 
consistent with estimates of {\it lower bounds} as discussed in 
\cite{ADD3}, for example.}.

For completeness let us note that the interaction cross section with CMB
photons is completely negligible because the cm-energy is about $10^{-13}$
that of a proton target.  On the other hand, the mean free path of
$UHE$ neutrinos with $E_\nu > 10^{19}$ eV and
$mb$-scale cross sections is of the order of a few kilometers in the upper
atmosphere.  This is still much larger than the mean free path of other
particles such as protons at the same energy.  If the ultrahigh energy
showers are indeed initiated by neutrinos, the $log(s)$ and $s^1$ models
predict considerable vertical spread in the point of origin of the shower.
The $s^2$ model is capable of much larger $\sigma_{tot}$.

Experimental extraction of cross sections is
complicated by shower development fluctuations, but it would be useful to
make all possible efforts to bound the mean free path of the primaries
responsible for $GZK$-violating events.  There is a further, interesting
variant of the ``double-bang" $\nu_{\tau}$ signal \cite{LP}: 
secondary events may come from a neutrino
dumping part of its energy in a first collision, undergoing a second
collision a few kilometers later.  This process may serve to separate
primaries with $mb$-scale cross sections from protons or nuclei,
which have much shorter mean free paths.

\subsection{Signatures of Massive spin-2 Exchange at Km-Scale Neutrino
Detectors}

It is also interesting to consider alternate signatures of the $\nu-N$
neutral-current cross section exceeding the Standard Model.  The cosmic
neutrino detectors, AMANDA \cite{amanda} and RICE \cite{rice} for example,
are expected to explore
the TeV-10 PeV energy regime.  The cross sections we find exceeding those
of the Standard Model might be tested in these experiments.
Of particular interest is the angular distribution of events.
The diffuse background cosmic ray neutrino flux is expected to be
isotropic.  We can, therefore, measure the deviations of $\sigma_{tot}$
from $SM$ predictions by measuring the ratio of upward- to downward-moving
events.  This ratio is plotted in Fig. 2 as a function of the incident
neutrino energy.  The plots show a few representative choices of $M_S$ and
extrapolation parameter $\beta$, using the $s^1$ model for illustration.
  The $up/down$ ratio starts to deviate very
strongly from the $SM$ value for an incident neutrino energy greater than
about 5 PeV. Beyond about 5 PeV the ratio falls very sharply to zero.  This
in principle can be measured at RICE \cite{rice}, which is sensitive to
precisely the
energy regime of 100 TeV to 100 PeV. The event rate in this energy region
is also expected to be significant.
Note that only the neutral-current
events are affected.  A more detailed, but also attractive extension of
this technique is $UHE$ Earth tomography, recently shown practicable with
existing flux estimates, and also capable of measuring $\sigma_{tot}$
indirectly \cite{tomo}.   The graviton-exchange predictions are not
especially sensitive to the precise value of $\beta$ in this region, but do
depend strongly on the cutoff scale $M_S$.  Large scale detectors such as
RICE will be able to explore the range of $M_S< 2$ TeV.

We conclude by reiterating that the highest energy cosmic rays are on the
verge of confronting fundamental particle physics.  Exciting projects
underway, including AGASA, Hi-RES, and AUGER should be able to collect
enough data to resolve the issues.  The puzzle of $GZK$
violating events can be experimentally resolved by establishing neutral
primaries via angular correlations.  That feature would imply new physics
of neutrinos, with the cross section then indicating massive spin-2
exchange rather directly. Experimentally bounding the primary interaction
cross section below that of protons is another viable strategy in the $s^1$
and $log(s)$ scenarios. 

 Current limits on $KK$ modes of extra dimension
models predict sufficiently large $\nu-N$ cross sections to produce such
interactions, and are consistent with observation of events in the upper
atmosphere at primary energies greater than $10^{20}-10^{21}$ eV.
Depending on the observational outcome, then, the subject matter could
become very important, and further development is well warranted.

{\bf Acknowledgments:} We thank Tom Weiler for discussions and suggestions
on the manuscript. PJ thanks Prashanta Das, Prakash Mathews and Sreerup
Raychaudhuri for useful discussions. 
This work was supported in part by U.S. DOE Grant number
DE-FG03-98ER41079 
and the {\it Kansas Institute for Theoretical and Computational Science}.

\newpage

\begin{figure}
\psfig{file=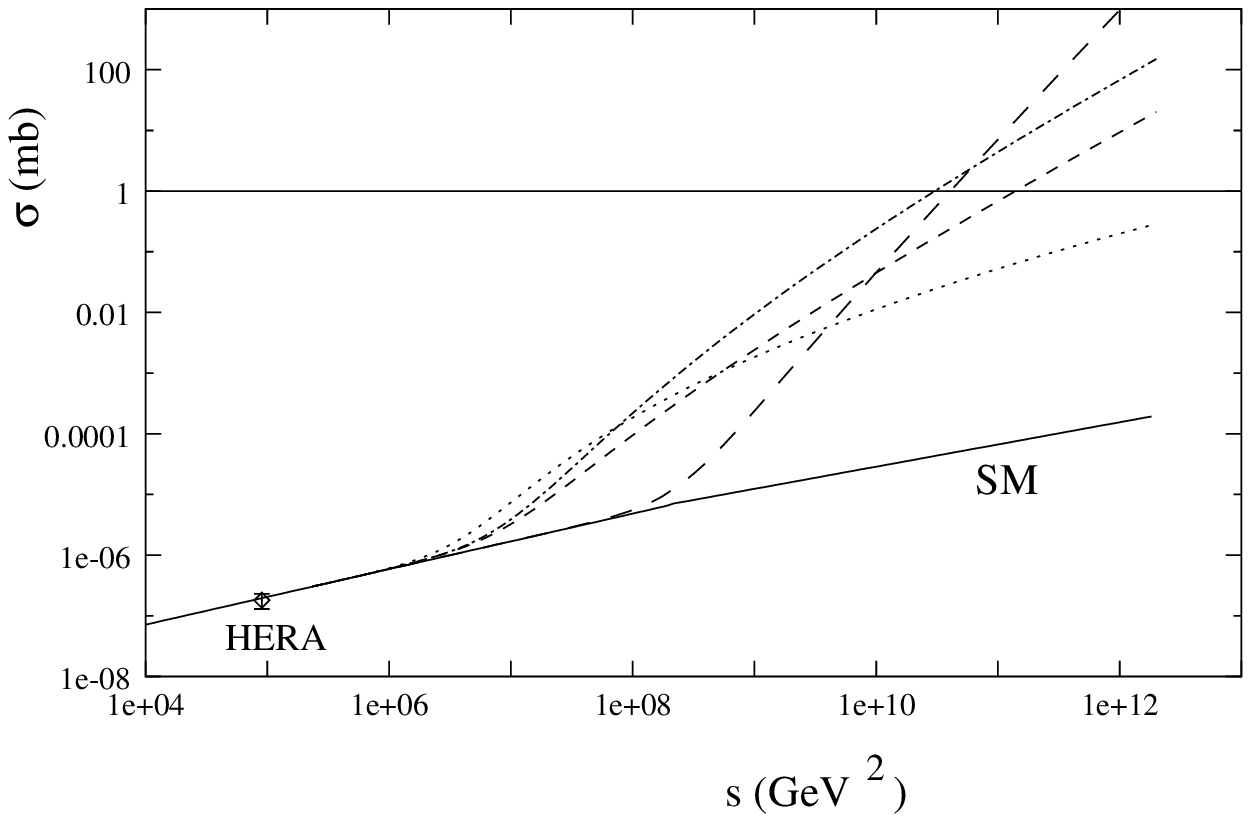}
\caption{The $\nu N$ cross section in the Standard Model ($SM$)
compared to a theory with large extra dimensions and three different
models for the unitarity extrapolation between perturbative to non-
perturbative regimes.  The dotted line shows the $log(s)$ growth case with
$M_{S}$ = $1$ TeV and $\xi$ = 10.  The short dashed and dash-dotted lines
show $s^1$ growth with $M_{S} = 1$ TeV and $\beta = 1$ and 0.1 
respectively.  The long
dashed line shows $s^2$ growth with $M_{S} = 3$ TeV and $\beta = 1$.  The
contribution from
massive graviton exchange is negligible at low energies but rises
above the $SM$ contribution when $\sqrt s > M_{S}$, reaching typical
 hadronic cross sections
at incident neutrino energies in the range $5\times10^{19}$ to
 $5\times10^{20}$ GeV.
The HERA data point is shown for comparison. 
The approximate minimum value required for $\nu N$ cross-section,
 $\sigma=1$ mb, is indicated by the horizontal straight line.} 
\end{figure}

\begin{figure}
\psfig{file=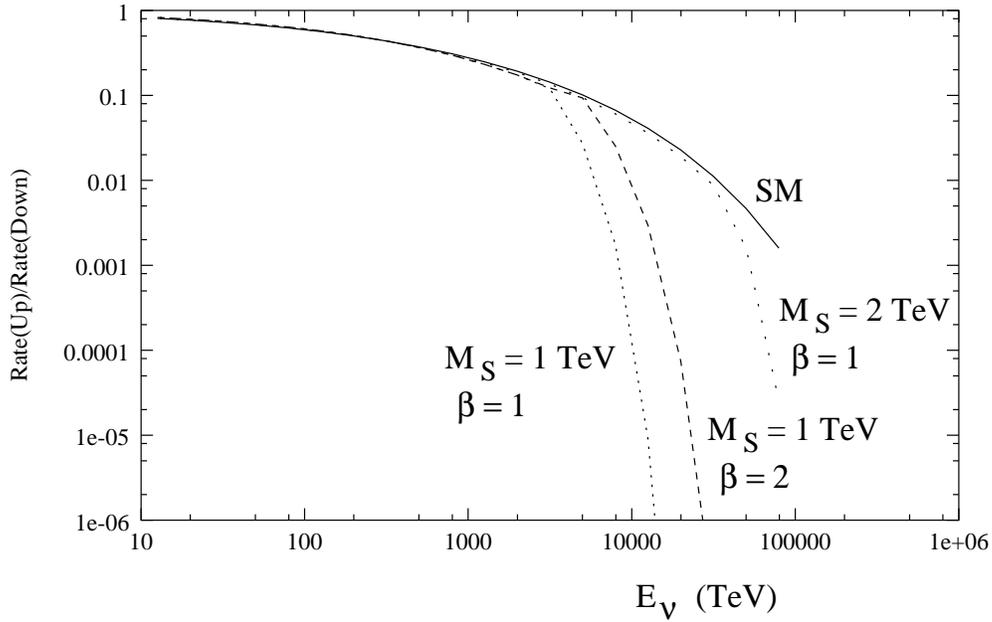}
\caption{The ratio of upward- to downward-moving events at $UHE$ cosmic ray
neutrino detectors as a function of
the incident neutrino energy. The solid curve corresponds to
the $SM$ prediction. Other curves include the contribution due to low scale
gravity models with $s^1$ cross section growth for
several different representative choices of the cutoff parameter
$M_S$ and the dimensionless extrapolation parameter $\beta=1$. Since the
extrapolation is small, the result is not especially sensitive to the
precise value of $\beta$.
}
\end{figure}
\end{document}